\documentclass[10pt,technote]{IEEEtran}
\usepackage{graphicx}
\usepackage{subfig}
\usepackage{cite}
\usepackage{url}

\begin{document}


 \twocolumn[%
 \centerline{\Large \bf A Literature Survey of Cooperative Caching in Content Distribution Networks} 

 \medskip
 \centerline{\bf Jing Zhang}
\centerline{Electrical and Computer Engineering}
\centerline{Drexel University}
\centerline{Email: jz334@drexel.edu}
 \centerline{May 10, 2012}
 \bigskip

\begin{abstract}
\boldmath{ Content distribution networks (CDNs) which serve to deliver web objects (e.g., documents, applications, music and video, etc.) have seen tremendous growth since its emergence. To minimize the retrieving delay experienced by a user with a request for a web object, caching strategies are often applied - contents are replicated at edges of the network which is closer to the user such that the network distance between the user and the object is reduced. In this literature survey, evolution of caching is studied. A recent research paper [15] in the field of large-scale caching for CDN was chosen to be the anchor paper which serves as a guide to the topic. Research studies after and relevant to the anchor paper are also analyzed to better evaluate the statements and results of the anchor paper and more importantly, to obtain an unbiased view of the large scale collaborate caching systems as a whole.}
\end{abstract}

 
\twocolumn]

\section{Introduction}
Decades ago when the World Wide Web was new, making an index of all the webpages was just like executing an Unix egrep command over 110,000 documents [1] . Within 7 years of its existence, the web had grown to compete with long existed information services such as television and telephone networks [5]. After around 20 years, in the month of May 2010 alone, U.S. Internet users watched nearly 34 billion videos [2]. Nowadays, omnipresent data access and data sharing for a large number of end-users are enabled by content distribution systems, such as on-demand video services [15], file-sharing networks [21], and content clouds [20].
\newline\indent To follow the exponential growth in demand of web service [8], caching was introduced to reduce the retrieve latency. Caching, however, was not a new idea at all. The first web browsers, such as Mosaic [9], were capable of caching web objects for later reference and thus reduce bandwidth for web traffic and latency to the users. Caching rapidly developed from being local for a single browser to being shared serving number of clients within a certain institution [8].
\newline\indent This literature survey is organized as follows: the evolution of caching is described in section II. In section III, a chosen anchor paper in the field of distributed caching is analyzed. Further research results in the field are presented in section IV. Conclusion and anticipated future work are discussed in section V.
\section{Evolution of Caching}
In 1995, Jacobson[3] proposed that caching could deal with the exponential growth of the internet. According to Jacobson, data has to find \emph{local} sources near consumers rather than always coming from the place it was originally produced. To match with the exponential growth of the internet, cache has to grow as fast as the internet - exponentially. The question remains as to how should we architect \emph{lots} of caches.
\newline\indent The taxonomy of caching was identified in 1982. Dowdy et al. proposed that an acceptable File Assignment Problem (FAP) solution assigns files to the nodes in some ``optimal" fashion. ``optimality'' was identified to be measured by cost and performance [4].
\subsection{Single Cache and Multiple Caches}
 In 1999, Shim et al. introduced a single cache algorithm in [5]. Albeit relatively simple with single cache proxy architecture, the LNC-R-W3-U algorithm took into consideration both cache replacement and consistency maintenance for web proxies. In 2000, Fan et al. described a multiple caching protocol - ``summary cache''. In the summary cache protocol, every proxy keeps a summary of the cache directory of each participating proxy and checks the summaries for potential hits before sending any queries [6]. Fan et al. recognized that to fully benefit from caching, caches should cooperate and serve each other's misses to reduce the total traffic through the bottleneck. 
\subsection{Hierarchical Caching}
	To increase the hit rate of web cache, large scale caching structure in which caches cooperate with each other were brought into sight - ``hierarchical'' and ``distributed'' caching. In hierarchical caching, caches are placed in different network levels [8]. As early as 1993, polynomial time algorithm for hierarchical placement problem was presented [12]. Harvest structure was considered to be the first hierarchical caching structure[8]. It was named to illustrate its initial focus on reaping the growing crop of Internet information [7]. Figure 1 is the overall Harvest architecture in which a Harvest Gatherer collects indexing information from across the internet, while a Broker collects information from many gatherers and provide query interface to the gathered information. Brokers can also collect information from other brokers to cascade the views from others [7].
\begin{figure}[!t]
\centering
\includegraphics[width=2.5in]{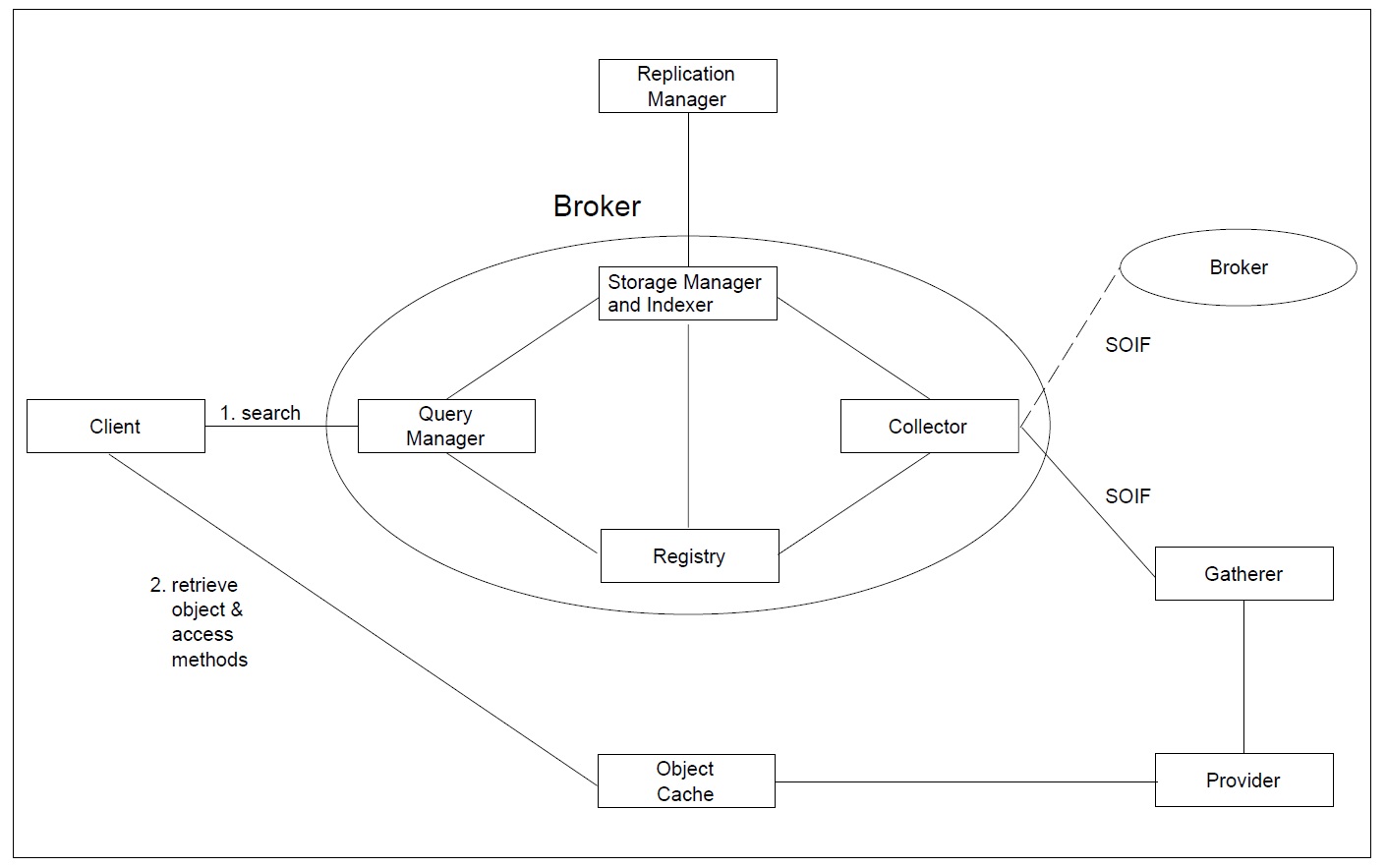}
\caption{Overall Harvest Architecture}
\label{fig_sim1}
\end{figure}
\newline\indent Advantages with hierarchical cache are reduced network distance to hit a document and reduced administrative concerns comparing to distributed architecture whereas some problems associated with hierarchical caching include: every level introduces latency [10], high level caches may experience long queuing delays and file placement redundancy that same document gets stored in different levels [8].

\subsection{Distributed Caching}
Another approach to implement large-scale cache is distributed caching. In distributed caching, only caches at the edge of the network cooperate to serve each other's misses. In the year of 2001, it was realized that distributed caching were becoming popular with the emerging of new applications that allow distributions of web pages, images, and music since distributed caching has lower transmission times than distributed caching due to the fact of most traffic flows  through less congested lower network levels[8]. Distributed cache also creates fair share of loads for the system and does not generate hot spots with high load. Nonetheless, distributed caching has longer connection times comparing to hierarchical caching [8]. Rodriguez et al [8] proposed that a hybrid scheme with optimal number of cooperating caches at each level could improve performance of hierarchical and distributed caching with reduced latency, load and bandwidth cost.
\subsection{Other Caching Structures}
A new ``transparent'' structure developed around 2002 was called \emph{en-route caching}. An en-route cache intercepts a client request that passes through it. If the requested object is in the cache, object will be sent to the client and the request will no longer be propagating further along the path. Otherwise, the node will forward the request along the regular routing path [11]. 

En-route caching has several advantages: 1) it is transparent to both content server and clients 2) no request is detoured off the regular path which minimizes network delay for cache miss and eliminates extra overhead such as broadcast queries [11].

In more recent research, more specific problems of distributed caching were studied such as adaptive distributed cache update algorithm for mobile ad-hoc networks [13] and distributed selfish caching [14] which more realistically models content delivery networks (CDNs), and peer-to-peer(P2P) systems. In Laoutaris et al's study [14], traditional grouping of distributed resources to ensure scalability and efficiency is deemed to be common strategic goal. New classes of network applications (e.g., overlay networks and P2P) are more ``ad hoc'' in the sense that it is not as strictly dictated by organizational boundaries or ``strategic goals'' described above. Individual nodes were described to be autonomous in the sense that membership within the group is solely motivated by benefiting from the group and thus a fluid nature of group membership was identified to be expected. Focusing on the susceptibility of nodes being mistreated (i.e. mistreatment due to interactions between group members and use of common scheme for cache management for all members of the group), [14] analyzed causes of mistreatments and suggested approach for an individual node to decide autonomously whether to stay or leave a group. This study carries great significance since it models realistic scenarios and suggested decisions to take while encountering the problem.
\section{Anchor Paper}
Published in 2010, the anchor paper focuses on minimizing the bandwidth cost instead of focusing on minimizing retrieving latency in previous studies such as [8],[13]. Borst et al. chose to focus on bandwidth minimization seeing the great momentum fueled by popularity of Video on Demand libraries. For example, Youtube is estimated to attract tens of millions of viewers a day generating around 2000 TB of traffic [15]. Borst et al. reasoned that for high-definition videos with sizes of few GBs and hours-long durations, minimize bandwidth usage is of far more significance than reducing initial play-out delay by several hundred of milliseconds. Objectives of the research were defined as designing light-weight cooperative content placement algorithms to maximize the traffic volume served from cache and minimizing the bandwidth cost. Cache replication strategies were proposed as linear programming formulations and studied numerically. Performance of the light-weight cooperative cache algorithm was also guaranteed to be within a constant factor from global optimal performance in [15].

In this study, instead of focusing on general network topologies, Borst et al. opted to focus on specific topologies motivated by real system deployments. Figure 2 is the cache cluster model considered in the anchor paper. The cache cluster model need not be stand-alone but rather be a part of a larger hierarchical tree network as in Figure 3. The structure is mostly hierarchical which also has some degree of connectivity among nodes of the same level. This structure should fit into the category of hybrid cache structure described in section II since hierarchy and distribution caching coexist in it.
\begin{figure}[!t]
\centering
\includegraphics[width=2.5in]{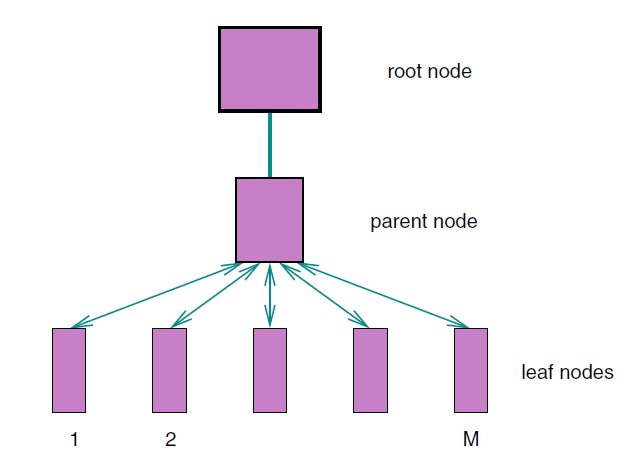}
\caption{Graphical Illustration of Cache Cluster}
\label{fig_sim2}
\end{figure}

\begin{figure}[!t]
\centering
\includegraphics[width=2.5in]{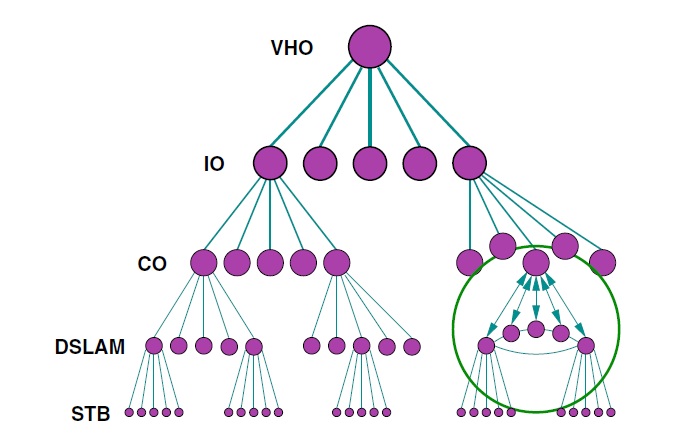}
\caption{Cache Cluster Embedded in Hierchical Tree Network}
\label{fig_sim3}
\end{figure}
Assuming equal bandwidth cost and cache sizes which is usually the way IPTV is configured [15], the bandwidth minimalization $P_{min}$ or cache maximum utilization $P_{max}$ could be modeled as
\begin{equation}
\label{eqn_example}
max    \sum_{n=1}^N s_nd_n(c^{```}p_n+c^`(M-1)q_n+Mc_0x_{0n})
\end{equation}
\begin{equation}
\label{eqn_example}
sub     \sum_{n=1}^N s_nx_{0n} \le B_0 
\end{equation}
\begin{equation}
\label{eqn_example}
 \sum_{n=1}^N s_n(p_n+(M-1)q_n) \le MB
\end{equation}
\begin{equation}
\label{eqn_example}
p_n+x_0n \le 1, n=1,...,N
\end{equation}
\begin{equation}
\label{eqn_example}
q_n+x_0n \le 1, n=1,...,N
\end{equation}
\\where $M$ denotes the number of `leaf' nodes, $s_1,...,s_N$ denotes the collection of $N$ content items, $d_{in}$ denotes the demand for $n$-th content item in node $i$, $c_{ij}$ denotes the cost associated with tranferring content from node $i$ to node $j$, $x_{ijn}$ is a boolean indicates whether requests for the $n$-th item at node $i$ are served from node $j$ or not.
\newline\indent The problem was simplified to a linear programming problem - \emph{ knapsack-type problem}. The problem was further analyzed in two scenarios - intra level cache cooperation and inter level cache cooperation. In intra-level cache cooperation case, the knapsack-type problem was further simplified to a knapsack of size $MB$ and $2N$ items of sizes $a_n=s_n, a_{N+n}=(M-1)s_n, n=1..N$. Based on the above insight, a  \emph{ Local - Greedy} algorithm was proposed. Borst et al stated that under symmetric demands, the worst case performance of { Local - Greedy} is $3/4$ of the global optimal performance. In the inter-level cache cooperation scenario, the solution can be even more simplified to a simple \emph{Greedy} algorithm. 
\begin{figure}[!t]
\centering
\includegraphics[width=3in]{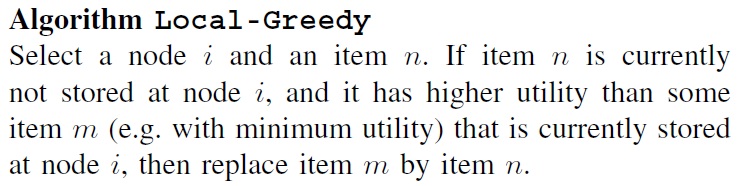}
\end{figure}
\newline\indent The demand was estimated by a Zipf-Mandelbrot distribution with shape parameter $\alpha$ and shift parameter $q$. Using the global optimum solution, the bandwidth costs were observed to be decreasing with increasing values of $\alpha$ which reflects the fact that cache performance improves as popularity function gets steeper as in Figure 4. However, it could be observed that neither full nor zero replication performs well across range of all $\alpha$ values. The author proposed that adjusting the degree of replication should be done according to the particular distribution and the \emph{ Local - Greedy algorithm} does just that. Therefore, the \emph{ Local - Greedy algorithm} is not only good in a way that it has low computing complexity but also that it combats the unknown popularity distribution problem. Further simulations on the \emph{ Local - Greedy algorithm} also showed that it performs close to the global optimum with a constant margin. When aging is slow, the algorithm converges fast (in around 2500 requests)to the global optimum.
\newline\indent From the performance analysis, we could see that \emph{ Local - Greedy algorithm} performs fairly well comparing to the global optimum solution. The worst-case performance ratio with respect to global optimum performance was guaranteed. The most fascinating feature of the algorithm is that instead of requiring prior knowledge of popularity distribution as needed by numerical analysis, the algorithm combats the problem of unknown popularity distribution in its nature of making locally optimal choices in every stage.
\begin{figure}[!t]
\centering
\includegraphics[width=2.5in]{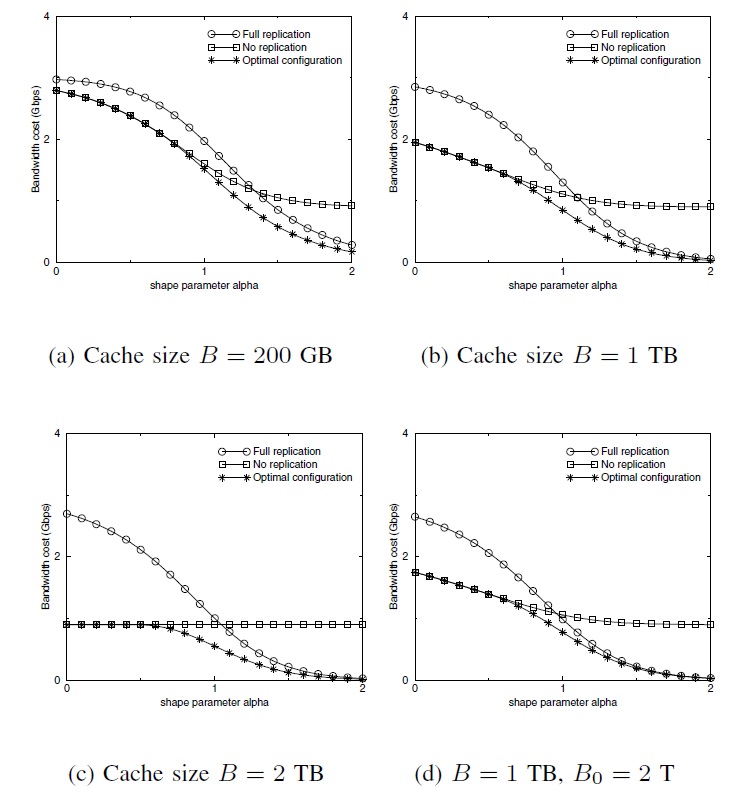}
\caption{Cache Cluster Embedded in Hierchical Tree Network}
\label{fig_sim4}
\end{figure}
\section{Further Researches}
The rapid increase of content delivery over the Internet has led to the proliferation of content distribution networks (CDNs) [18] which complied with Borst et al’s prediction in [15] and thus confirmed the significance of studying large scale cooperative caching system. Several research studies relevant to the anchor paper published after [15] are analyzed in this section.

Wireless access technology including Wi-Fi, HSPA+, 4G LTE have made broadband wireless connections (e.g., peak downlink rate of 100 Mbps in LTE) a near-term reality [16]. This advancement enabled user to stream videos on their hand-held devices. According to Dai et el, cache servers of Wireless Service Providers (WSPs) are typically deployed at Mobile Switching Centers (MSC) to locate video content closer to end users in order to maximize bandwidth efficiency. However, the dynamics of mobile users made resources provisioning at cache servers a great challenge. [16] was the first to suggest a collaborative caching mechanism between multiple WSPs while guaranteeing both fairness and truthfulness. The collaborative caching problem was formulated as a Vickrey-Clarke-Groves( VCG) auctions problem in game theory which encourages cache servers to cooperate for trading their bandwidth as commodities in auctions. Dai et al. suggested that the VCG auction from game theory fits for designing the mechanism for several reasons: buyers have to \emph{pay} for each successful trade which serves as incentives to contribute, \emph{truthfulness} can be ensured by the payment method in VCG auction in which buyers are willing to truthfully reveal their bidding information in an auction [16]. Simulation results shows that with maximizing social welfare in VCG auction problem the video streaming quality can be tremendously improved through the collaboration of caching among various WSPs.

One of the significances of [16] is that by acknowledging the benefits from collaborative caching in [15], it takes one step further considering how collaboration between several WSPs could improve the problem caused by dynamic nature of mobile users.

Similar to [16] in the consideration that nodes can make autonomous selfish decisions, [17] proposed a distributed algorithm which accounts for the selfishness of autonomous nodes and the churn phenomena (i.e. random ``join'' and ``leave'' events of nodes in the group). Although the algorithm proposed in [17] might not converge at equilibrium, it is still highly desired for several reasons: it decreases access cost for all nodes compared to greedy local or churn-unaware strategy, and it provides a fairer treatment to nodes according to their reliability while churn-unaware strategy can have higher overheads accessing unreliable nodes [17]. This study also went further beyond the study of the anchor paper [15] in a realistic consideration that nodes are becoming more dynamic recently with the advancement of mobile technology.

In [18], policies for request routing, content placement and content eviction for CDN were designed with the goal of small user delays. While [15] focuses on minimizing bandwidth usage considering video streaming scenarios, [18] focuses on achieving small user delay considering smart-phone applications, music and video files and concluded that low delays would be desired in every case. Although content in a cache that is close to a user is likely to experience shorter delay, placing massive contents at the nearest cache might be counterproductive since link capacity between cache and end-user is not infinite [18]. Figure 5 from [18] is an abstract switch model of a CDN network. 
\begin{figure}[!t]
\centering
\includegraphics[width=2.5in]{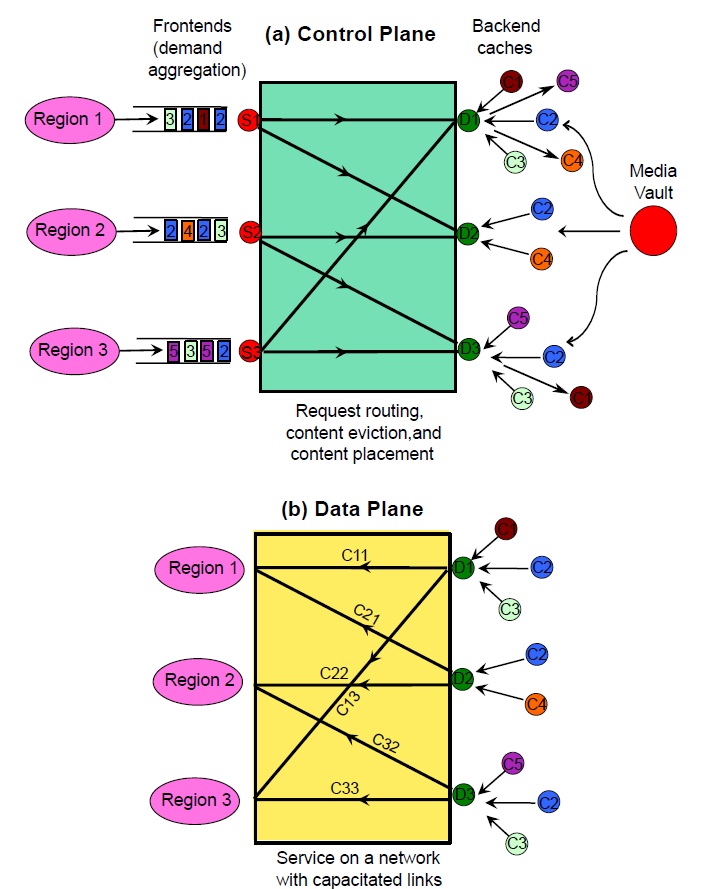}
\caption{A Content Distribution Network. (a) Control Plane: Requests arrive at
frontend servers (S), and must be routed to one of (possibly) several backend
caches (D) that have the content. Caches can only host a finite number of
content files (C), and the caches may be refreshed by placement and eviction
of content, (b) Data Plane: Content is served to end-users across a network
consisting of finite capacity links.}
\label{fig_sim}
\end{figure}
\newline\indent For a CDN as in Figure 5, each query could be potentially served from multiple backend caches and each frontend takes a decision on picking a backend cache. The constraint of the system is that the network connecting the backend cache has finite capacity, each backend cache host finite amount of content and refreshing content in the caches risks to incur a cost. Objective of [18] is to develop algorithms for jointly solving the request routing and content caching problems. Four algorithms satisfy the objectives were proposed.
\subsubsection{Periodic Max-Weight algorithm with random eviction}
The algorithm tries to stabilize a system using a Lyapunov function that is quadratic in queue lengths. [18] showed that algorithm is throughput optimal, and has bounded queue lengths which is desirable
\subsubsection{Periodic Max-Weight Scheduling with Min-Weight Eviction policy}
Queue size is related to the drift of Lyapunov function that larger negative drift of Lyapunov function implies shorter queue length which is the insight of the algorithm. This algorithm is also throughput optimal and has low computational complexity.
\subsubsection{Iterative Periodic Max-Weight algorithm}
Both 1) and 2) are throughput optimal but inefficient in link capacity usage. This algorithm attempts to maximize link capacity utilization. Amble et al showed that this is also a throughput algorithm and likely to have shorter queue length than PMW.
\subsubsection{IPMW Scheduling with Min-Weight algorithm}
The algorithm is an IPMW version of algorithm 2 that is also throughput optimal and has average queue length at most that of PMW with min-weight evictions.

The performance of the four algorithms proved that algorithms generating large, negative Lyapunov drift in a system are desirable since it indicates short average queue delays [18]. This study took a more ``network'' perspective model of content distribution networks by modeling it as a switch structure. The anchor paper [15] provides general analysis of large scale content distribution network while [18] takes a closer view to the network recognizing that capacity links are not infinite and tries to reduce content retrieving latency by minimize queuing delay of the abstract CDN model.

Bj¨orkqvist et al. proposed two Peer Aware Content Caching(PACC) for CDN networks in [19]. Similar to the anchor paper [15], objectives of [19] were recognized as minimize total content retrieval cost and optimize bandwidth consumption. As opposed to a specific topology setup in [15], [19] aimed on \emph{generic} three layer CDN systems where both vertical and horizontal retrieval is enabled. The two PACC policies proposed are PACC-AR and PACC-CL. PACC-AR is implemented distributively whereas PACC-CL collaborates with peer nodes in caching process. Simulation result shows that both policies come close to achieving the optimum diffusion and low retrieval costs especially the collaborated PACC-CL. Bj¨orkqvist et al argued that [15] is not applicable while evaluating from a system perspective(i.e., dimensions of the system size and storage capacity) and further stated that the framework in [19] is capable of evaluating the content retrieval costs. Further work suggested by [19] is to conduct experimental analysis. 
\section{Conclusion}
In this literature survey, the evolution of caching was studied as background information to better understand the current collaborative caching structures such as Content Distributed Networks (CDN). An anchor paper [15] focusing on large scale collaborating caching was analyzed. In the anchor paper, the total bandwidth minimization problem was reduced to a linear programming knapsack-type problem and further simplified to a \emph{ Local-Greedy algorithm}. The anchor paper successfully stressed the fact that collaborated caching improves bandwidth usage. In later on studies, [16] suggested a collaborative caching mechanism between multiple Wireless Service Providers (WSPs) to optimize caching performance for mobile users.  [17] focuses on developing a distributed caching algorithm that takes selfishness and churn phenomena(i.e., random ``join'' and ``leave'' in the group)  into consideration also due to the realistic fact that nodes are becoming more dynamic lately. [18] modeled CDN as a switch model and take into account finite link capacity between the cache and end-user which provides a more ``network'' perspective to the CDN than [15]. [19] proposed two Peer Aware Content Caching (PACC) policies for CDN networks also taken into account dimension of the system size and storage capacity as one step further from [15]. All the later on work suggested that nodes are being more and more mobile and more specific system models considering the system dimension should be developed. [19] suggested that experimental results should be done in its future work which is also an anticipation for all the work done currently in the field. With all the numerical analysis and simulation experiment work done in the cooperative caching field, it is time to examine the theory results by practical experiments.
\section{References}

\noindent [1] O.A. McBryan (1994). GENVL and WWW: Tools for Taming the. Web, Proc. First Int'l World Wide Web Conf., Elsevier, New. York, 79-90.
\newline [2] \url{http://www.comscore.com/Press_Events/Press_Releases/2010/6/comScore_Releases_May_2010_U.S._Online_Video_Rankings}
\newline [3] V. Jacobson (1995). How to kill the internet, presented at SIGCOMM'95 Middleware Workshop, [Online]. Available: \url{ftp://ftp.ee.lhl .gov/talks/vj -webflame.ps.Z}
\newline [4] L.W. Dowdy, D.V. Foster (1982). Comparative models of the ﬁle assignment problem. ACM Comput. Surv 14, 287–313.
\newline [5] J. Shim , P. Scheuermann , R. Vingralek (1999). Proxy Cache Algorithms: Design, Implementation, and Performance, IEEE Transactions on Knowledge and Data Engineering, v.11 n.4, p.549-562. 
\newline [6] L. Fan, P. Cao, J. Almeida, A.Z. Broder (1998). Summary cache: A scalable wide-area Web cache sharing protocol. In: Proc. ACM SIGCOMM ’98, 254–265.
\newline [7] The Harvest Group. (1994) Harvest Information Discovery and Access System. [Online]. Available: \url{http://excalibur.usc.edu}
\newline [8] P. Rodriguez , C. Spanner , E.W. Biersack (2001). Analysis of web caching architectures: hierarchical and distributed caching, IEEE/ACM Transactions on Networking (TON), v.9 n.4, p.404-418.
\newline [9] K. Claffy, H. W. Braun (1994). “Web traffic characterization: An assessment of the impact of caching documents from NCSA’s web server,” in Electronic Proc. 2nd World Wide Web Conf.’94: Mosaic and the Web.
\newline [10] A. Chankhunthod et al. (1996). A hierarchical internet object cache. USENIX Technical Conf., San Diego, CA.
\newline [11] X. Tang , S.T. Chanson (2002). Coordinated En-Route Web Caching, IEEE Transactions on Computers, v.51 n.6, p.595-607
\newline [12] A. Leff , P.S. Yu , J.L. Wolf (1991). Policies for efficient memory utilization in a remote caching architecture, Proceedings of the first international conference on Parallel and distributed information systems, p.198-209, Miami, Florida, United States
\newline [13]X. Yu and Z. M. Kedem (2005). A distributed adaptive cache update algorithm for the dynamic source routing protocol. In IEEE INFOCOM'05, Miami, FL, USA.
\newline [14] N. Laoutaris, G. Smaragdakis, A. Bestavros, I. Matta, and I. Stavrakakis (2007). Distributed selfish caching. IEEE Transactions on Parallel and Distributed Systems, 18:1361–1376.
\newline [15]S. Borst, V. Gupta, and A. Walid (2010). Distributed caching algorithms for content distribution networks. In Proc. IEEE INFOCOM, San Deigo, CA.
\newline [16]J. Dai, B. Li, F.-M. Liu, B. Li and J.-C. Liu (2012). "Collaborative Caching in Wireless Video Streaming Through Resource Auctions," accepted and to appear in IEEE Journal on Selected Areas in Communications, Special Issue on Collaborative Networking Challenges and Applications.
\newline [17] E. Jaho, I. Koukoutsidis, O. Stavrakakis, and I. Jaho (2012). Cooperative content replication in networks with autonomous nodes. Comput. Commun. 35, 5, 637-647.
\newline [18] M. M. Amble, P. Parag, S. Shakkottai, and L. Ying (2011). “Content-Aware Caching and Traffic Management in Content Distribution Networks,” IEEE INFOCOM 2011, Shanghai, China.
\newline [19] Mathias Björkqvist, Lydia Y. Chen, Xi Zhang (2011). Minimizing Retrieval Cost of Multi-Layer Content Distribution Systems. ICC 2011: 1-6
\newline [20] \url{http://www.cloudbook.net/reservoir-gov.}
\newline [21] \url{http://wua.la.}
\end{document}